# The evolution of group differences in changing environments

Arbel Harpak[1,*] & Molly Przeworski[1,2,*]

[1] Department of Biological Sciences, Columbia University
[2] Department of Systems Biology, Columbia University
[*] To whom correspondence should be addressed: ah3586@columbia.edu or mp3284@columbia.edu

## Abstract

The selection pressures that have shaped the evolution of complex traits in humans remain largely unknown, and in some contexts highly contentious, perhaps above all where they concern mean trait differences among groups. To date, the discussion has focused on whether such group differences have any genetic basis, and if so, whether they are without fitness consequences and arose via random genetic drift, or whether they were driven by selection for different trait optima in different environments. Here, we highlight a plausible alternative, that many complex traits evolve under stabilizing selection in the face of shifting environmental effects. Under this scenario, there will be rapid evolution at the loci that contribute to trait variation, *even when the trait optimum remains the same*. These considerations underscore the strong assumptions about environmental effects that are required in ascribing trait differences among groups to genetic differences.

## Introduction

The last couple of decades of research in human genetics have substantiated the perspective on phenotypes that was developed by quantitative geneticists over the past century[1,2]. In particular, it is now clear that most human traits of interest, whether morphological, physiological or behavioral, are "complex", meaning that individuals differ in their heritable phenotypes because of numerous small contributions of loci scattered throughout the genome and varying environments[3–6]. A paradigmatic example is height [7,8]. The height of an individual adult depends on myriad alleles they carry in their genome, features of the environments in which they develop, grow up and live, and potentially intricate interactions between their genetics and the environment.

Given the advent of genome-wide association studies (GWAS), it is now possible to map loci that contribute to human trait variation and to construct predictors of individual trait values, so-called "polygenic scores" (PGS)[9,10]. Although in their infancy, PGS have already been deployed for many purposes, notably in disease prognosis (e.g. refs. 11–14). PGS have also been used in comparisons between sets of individuals with different genetic ancestries (e.g., refs. 15–20). As numerous authors have shown, the construction of PGS is fraught with difficulties, because of uncertainty about which genetic variants truly have a non-zero effect on the trait (i.e., are "causal"), population structure confounding and many other factors[21–26]. Partly as a result, even where existing PGS explain a substantial proportion of trait variation in the GWAS sample, they are poor predictors of trait values in individuals that differ from the GWAS sample in their genetic ancestry (or in other characteristics)[21,22,27,28]. Given these limitations, it is no surprise that PGS distributions differ among genetic ancestry groups and indeed current comparisons of PGS are very difficult to interpret[29].

An underlying question remains, however: Under what conditions should we expect the true, heritable component of trait variation to differ among groups and what, if anything, can we learn from genetic differences about the selection pressures that shaped trait differences? To date, the question of how genetic differences between groups might arise has been considered primarily in light of two possibilities: that a trait has no fitness effects (i.e., is "neutral") or that the trait optimum changed recently, leading to directional natural selection. It is well understood that if a focal trait and correlated traits do not affect fitness, then the frequencies of alleles that contribute to variation in the trait will drift over time and consequently vary somewhat across the globe[19,30]. Similarly, it is intuitive that if a complex trait is under directional selection, such that higher (or lower) trait values are favored, allele frequencies will change more rapidly than under drift alone[19,31–38]. In these cases, it is also well appreciated that because environmental effects differ over time and across the world, the resulting differences in trait values between groups are unpredictable (e.g., refs. 29,30,39,40).

In this Perspective, we review these results but emphasize an alternative scenario, which may be quite common[31]: that the trait is under ongoing stabilizing selection, i.e., that there exists a stable optimal trait value and selection against values far from it. As we outline, under stabilizing selection for a fixed trait optimum, shifting environmental effects on the trait will lead to rapid changes in frequency at loci that influence trait variation. In other words, there will be transient polygenic adaptation even in the absence of a change in trait optimum. We discuss the implications in two settings where genetic differences have been interpreted in terms of trait differences: comparisons of human groups and tests for polygenic adaptation.

**Assumptions**

For concreteness, we focus on two geographically-separated groups of individuals (henceforth "populations") who derive from a common ancestral group and ignore demographic effects such as those induced by changing population sizes or structure within populations. In practice, the classification of individuals into groups is at the discretion of the researcher and may depart from our assumptions (for example, because the groups subsequently came into contact and intermixed[41]). Even this simplified model suffices to highlight many interpretative challenges, however.

We consider a quantitative trait, whose individual values ($Y$) are given by the sum of the total genetic effect ($G$) across all causal sites in the genome, polymorphic or not, and the total environmental effect ($E$),
$$Y = G + E. \qquad \text{(eq. 1)}$$
Here, the genetic differences among individuals are due to many independently-evolving loci of small effect. We make the standard assumption that alleles are additive at a locus and across loci, thereby ignoring possible dominance or epistatic effects. We do so because an additive model provides a very good fit to existing GWAS data for humans[6,42,43]; because most theory for the evolution of complex traits has been developed for this model (e.g., refs. 19,26,38,44,45); and most importantly, because, while it is a simplification, the qualitative points that we make do not depend on it.

As is also standard, we assume that within each population, $E$ is Normally distributed and independent of $G$[7]. For now, we also ignore possible interactions between genetics and environment. We note that the word "environment" has ambiguous uses in evolutionary genetics. Often, it refers to the ecological context that shapes the fitness function (i.e., the relationship between values of $Y$ and fitness). Here, as is common in quantitative genetics, we mean the "environmental effect" on the trait, namely $E$ in eq. 1.

In what follows, we focus on the *true* genetic effects on the trait and not what can currently be learned about them. We refer to the true, total genetic effect $G$ on a trait as the "polygenic effect" (this quantity is akin to the breeding value in quantitative genetics[9] and to what in other contexts is called the genetic or genotypic value[2,19]). For a given individual, the polygenic effect is constructed by summing the alleles over every site in their genome, weighted by their effects on the trait value. Given this set-up, a PGS can be viewed as an estimate of the polygenic effect (shifted by a constant). As noted above, PGS rely on results from GWAS and currently suffer from numerous limitations (e.g., refs. 21,22,26,28). Unless otherwise stated, however, when we discuss the genetics of trait variation, we do not mean the PGS, but the polygenic effect, assuming knowledge of the exhaustive set of causal loci and their exact effects on the trait.

**Polygenic effects are expected to evolve under most plausible scenarios**

**A trait with no fitness effects**. If fitness is the same for all trait values and correlated traits are similarly neutral, the distribution of polygenic effects and trait values will be approximately Normal in each population, and the expectation of the polygenic effects will be the same. In any given realization of the evolutionary process, however, the mean polygenic effect in the two populations will diverge as allele frequencies undergo genetic drift[19,46,47]. In humans, the difference in means will tend to be small (relative to the variation within populations), as there has been little drift among populations[19], and either population may end up with a slightly higher mean polygenic effect, with equal probability (*Fig. 1, case 1*).

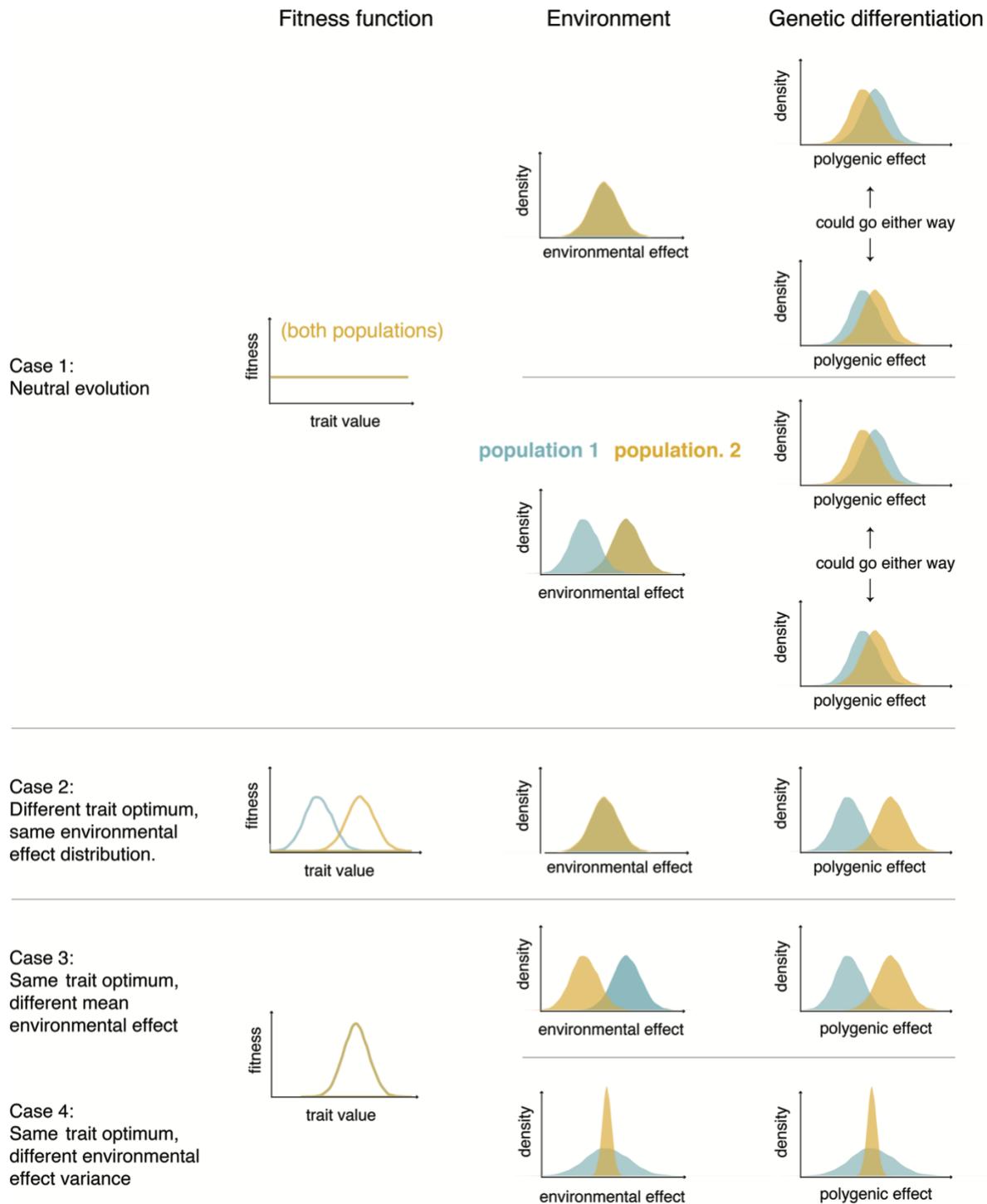

**Figure 1**

Random genetic drift is always occurring, so will contribute to differences in polygenic effects regardless of additional selection pressures. For that reason among others, a neutral model may be a useful point of departure. Yet it is unlikely to apply strictly to many traits. Indeed, the flip side of the high polygenicity of many traits is that variants in the human genome often affect more than one trait[6,48–51]. Thus, even when the focal trait itself has no effects on evolutionary fitness, variation in the trait may often be under selection because of the pleiotropic effects of the loci that shape it[31,52].

**Directional and stabilizing selection**. One alternative to strict neutrality is for the focal trait to have been under recent directional selection, such that higher (or lower) trait values were favored[6,19,35,53–56]. As an example, it has been proposed that darker skin pigmentation was beneficial to individuals that moved to a geographic location with higher levels of ultraviolet radiation (UVR)[57]. In this loose conceptualization of directional selection, higher trait values are always better, at least transiently.

Another formulation, which may be more realistic, is to consider a trait evolving under stabilizing selection, such that most individuals have trait values around an intermediate optimum, then imagine a sudden change in optimum (due, say, to a change in UVR levels). (In this case, a higher trait value is not always favored, for example because it may greatly exceed the new optimum value.) This scenario has been the focus of many decades of research in quantitative genetics[1,2,6,31,52,58] but has received less attention in human genetics (but see refs. 59–62).

Stabilizing selection can arise from trade-offs between the fitness advantages and disadvantages provided by the trait; a textbook example is birth weight, which is deleterious when too low or too high[63]. Importantly, stabilizing selection on a trait can also be only "apparent", arising from pleiotropic effects of variants contributing to the trait (ref. 31 and references therein). For example, if a focal trait (height) has no direct effect on evolutionary fitness, individuals with extreme phenotypes (very tall or very short) could be selected against because carrying too many alleles that nudge in the same direction (many tall alleles or many short ones) also tends to lead to other phenotypes that are deleterious[48,64]. Even if increased height itself were favored, its dynamics may still be dominated by stabilizing selection because of deleterious consequences (e.g., on cancer risk or musculoskeletal problems) of carrying too many height-increasing alleles (ref. 52 and references therein). Depending on one's view of what constitutes a trait, the distinction between direct and apparent stabilizing selection may be somewhat semantic[6,48].

For now, we focus on a single trait under stabilizing selection, in which case it is standard to model the polygenic effect and the trait as Normally distributed (assuming no mutation bias) (e.g., ref. 44), and consider the impact of changing the optimum, say to a higher trait value. Assuming this new optimum is not exceedingly far off (relative to the trait variance within the population), it is rapidly attained through a slight, average increase in the frequencies of many trait-increasing alleles[33,37,38,65]. The transient period of directional selection is followed by a much longer one during which stabilizing selection on the trait again dominates allele-frequency dynamics[35,38]. Regardless, there will be a rapid shift upward in the polygenic effect (*Fig. 1, case 2*).

**Stabilizing selection for a fixed optimum**. While possible changes in complex traits over the course of human evolution have received a lot of attention, there is often little reason to expect contemporary populations to differ in their fitness optima. Thus, an important alternative to consider is what happens when the trait is under ongoing stabilizing selection for the same optimum in both populations. If we assume the same fitness function and the same distribution of environmental effects on the trait in the two populations, then the distribution of the polygenic effects will also be the same.

For many traits, however, it is unrealistic to assume that the distribution of salient environmental effects is identical across groups or over time. As just one example, we know that environmental effects on height (such as diet) vary over time and among countries (e.g., ref. 66). More generally, differences in environmental effects between populations can arise from many sources, if individuals in the groups do not have identical distributions of diets, living conditions, medical care, incomes and so forth (e.g., refs. 22,67–69).

The seemingly innocuous observation that salient environmental effects probably differ somewhat between any two populations has a key implication. If we assume the same fitness function (including the same optimum) in the two populations, but posit that the environmental effects tend toward smaller trait values in population 2, then we expect selection favoring a higher polygenic effect in population 2 and a concomitant upward shift in the mean polygenic effect (*Fig. 1, case 3*). For instance, if we imagine poorer maternal nutrition on average in population 2 and the same optimal birth weight in both populations, then we might expect a higher polygenic effect in population 2—not because there is selection for higher birth weight in population 2, but because of selection to counteract the poorer maternal nutrition (i.e., "genetic compensation"; refs. 70–72). This type of compensation has been invoked to explain adaptation to high altitude hypoxia in mammals, the hypothesis being that genetic changes in high altitude populations counteracted the maladaptive (environmental) effects of acclimatization, thereby maintaining similar hematological profiles to low altitude populations[73]. As these examples illustrate, *there can be polygenic adaptation and a shift in mean polygenic effect between populations even when there is no difference in trait optimum*.

We may also consider a case in which the fitness functions are the same in the two populations but the variance of environmental effects differs, for instance because individuals in population 1 encounter more heterogeneous environments than those in population 2. In this case, the fitness difference between any two polygenic effects will be smaller in population 1 than in population 2 because fitness in population 1 depends more heavily on environmental effects[74]. (For simplicity, we ignore further complications such as canalization and selection for robustness[62,75–77].) Because the effects of stabilizing selection are more pronounced in population 2 than in population 1, at equilibrium, polygenic effects in population 2 will be less variable (*Fig. 1, case 4*, where for ease of interpretation, the mean environmental effects are the same).

If we instead suppose that the focal trait is not under direct but under apparent stabilizing selection, as a result of correlations with other traits, the effects of shifting environmental effects on polygenic effects are harder to predict; for one, they will depend on the precise nature of the pleiotropic effects[6,48,52,78]. But the qualitative point remains: in the face of changing environmental effects, stabilizing selection for a fixed optimum, like selection for a new trait optimum, can lead to polygenic adaptation.

To recap, the genetic contribution to variation in a trait will evolve whether the trait is strictly neutral, dominated by transient directional selection for a new optimum, or under stabilizing selection for the same optimum, as long as salient environmental effects differ among groups of people or over time. Thus, we should expect that mean polygenic effects will not be identical between populations.

**Interpretations of trait differences rely on strong assumptions about environmental effects**

Regardless of how population differences in polygenic effects arise, their mere existence implies nothing about their relationship to trait differences. Indeed, when both polygenic and environmental effects differ, all bets are off (see, e.g., refs. 29,30,39,71,72,79): The mean polygenic effect could differ substantially even when the mean trait value does not differ at all (Fig. 2, case A). The mean polygenic effect could be higher in population 2 when the mean trait value is lower (Fig. 2, case B). Or the differences in polygenic effect and trait values could align (Fig. 2, case D). Only when environmental effects on the trait are similar in the two populations, should we expect mean trait difference to mirror the difference in mean polygenic effect (Fig. 2, case C).

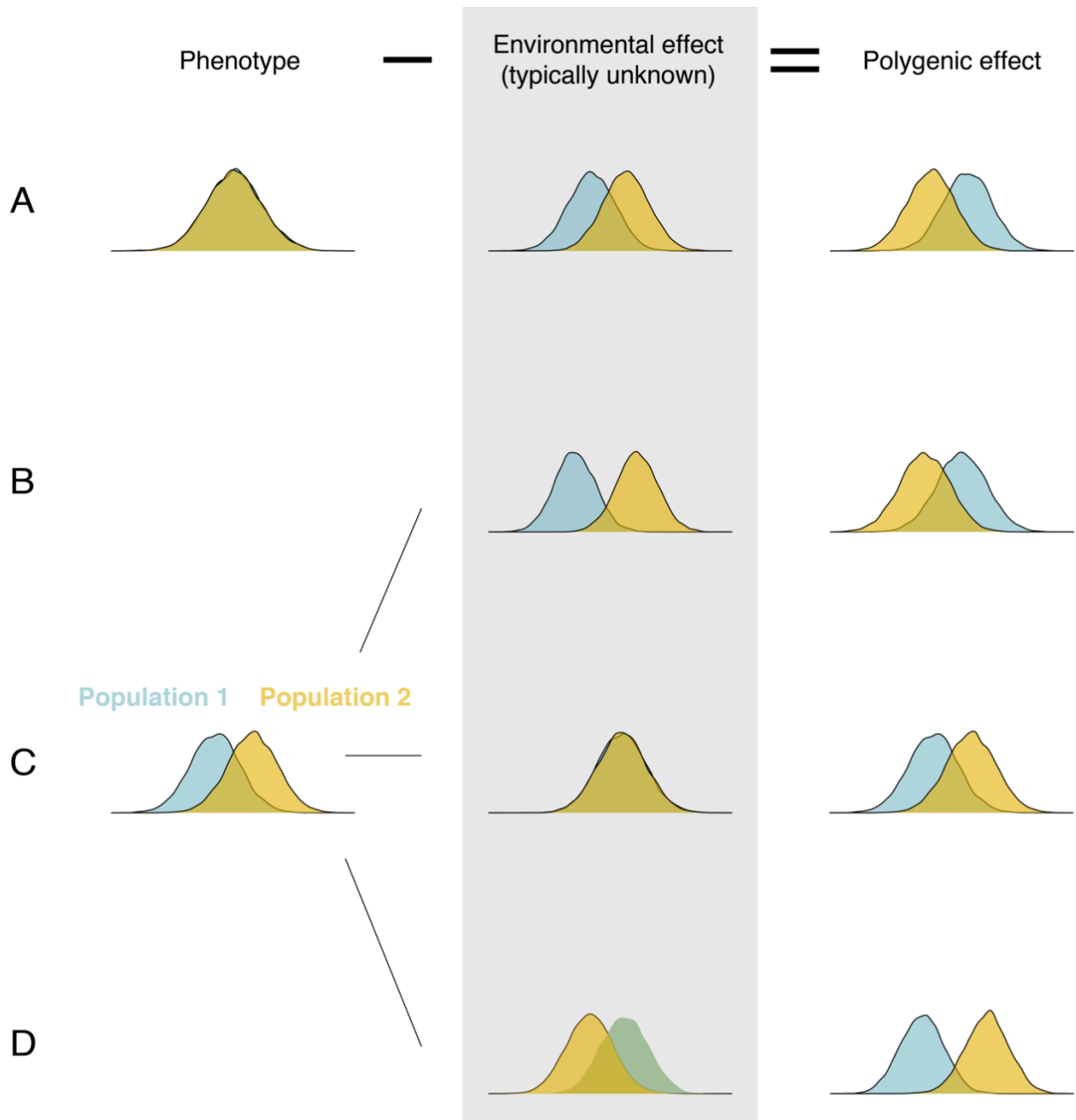

**Figure 2**

Such considerations highlight an important—and at times implicit—assumption made in comparisons of PGS and observed mean trait values across groups with different genetic ancestries: that environmental effects are identical (e.g., refs. 16,80,81). As noted in these studies, mean differences in PGS among groups often do not track differences in mean trait values. For example, on the basis of the PGS, one would predict that, on average, individuals in one group should be shorter than individuals in the other, when in fact they are taller. The assumption comes in when this observation is offered as evidence that PGS are unreliable outside of the ancestry in which they are estimated. There are many reasons to expect that PGS would not port reliably from a GWAS sample to individuals of other genetic ancestries and several lines of empirical evidence that they do not[16,21,22,25,26]. Even if all current limitations of PGS were surmounted, however, a discrepancy between a between-population difference in PGS and corresponding trait may well persist, owing to differences in environmental effects between populations.

**Adaptive differences in polygenic effects could reflect changes in trait optima or shifts in environmental effects**

Assumptions about environmental effects also matter critically when interpreting evidence for adaptive differences in polygenic effects. Under the standard quantitative-genetic assumptions, a shift in the optimum of a trait given a fixed environment is mathematically equivalent to shifting the mean environmental effect while keeping the optimum fixed. Thus, an increase in the polygenic effect for a trait such as height could reflect selection for increased height in a fixed nutritional environment or selection for the same optimal height in an environment where nutritional effects on height are suddenly much lower, or the combined effect of the two.

This equivalence between an environmental shift for a fixed optimum and a shift in the trait optimum shines a different light on the interpretation of tests for polygenic selection. Given mounting evidence that many adaptations of interest in human evolution were likely polygenic[6,35,53,54,82–84], there has been a recent push to identify the signature of polygenic adaptation in sets of loci associated with a trait in a GWAS[18–20,32,65,85–89]. Interpreting the results is extremely tricky, as a number of authors have underscored[19,28–30,40], both because of technical challenges (e.g., residual stratification in GWAS and the poor portability of PGS[21–23,28,81,90,91]) and deeper conceptual issues, such as the problem of selection on correlated traits[6,30,52,60,88,92,93]. Nonetheless, where these methods have identified possible signatures of polygenic selection, they have often been interpreted as pointing to a trait (or a correlated trait) whose optimum value has shifted in response to natural selection in the course of human evolution. An alternative interpretation is that the optimal trait value has remained the same, but environmental effects have shifted[6,19,40,72,87]. Thus, evidence of polygenic adaptation could be reflective of a change of trait optimum, but need not be. Of course, it is also possible that drastic changes in environmental effects are often accompanied by shifts in the optimum trait value.

This interpretative ambiguity is heightened by the fact that traits are inevitably somewhat arbitrary constructs[6,40,48]. Consider body shape as one example. Bergmann's rule proposes that in humans, as in other mammals, body shape varies across the globe in part in order to minimize heat dissipation. Under Bergmann's rule, greater body mass, and hence higher ratios of surface area to volume, are favored in colder climates (e.g., refs. 94,95 and references therein). This selection pressure could be interpreted either as favoring different body shapes in different ecological settings or as maintaining the same heat dissipation optimum in the face of a changing environmental effect. Another example may be provided by skin pigmentation levels, which mediate UVR penetration into the skin and vary with UVR exposures across the globe. The extent of UVR penetration is thought to be subject to a trade-off that arises from dual effects of UVR on folate degradation and vitamin D synthesis—two critically important vitamins[57]. Numerous studies have reported evidence for polygenic adaptation at the loci that contribute to variation in skin pigmentation levels, both across the globe and over time (e.g., refs. 96,97). These signatures of polygenic adaptation are usually viewed as resulting from repeated episodes of directional selection on skin pigmentation in environments with different UVR levels, but could also be seen as arising from selection for a similar degree of UVR penetration in the face of varying UVR levels. In other cases, much less is known about the relationship between the traits and fitness, and the interpretation will be all the more challenging.

That polygenic adaptation can arise from stabilizing selection for a fixed optimum in shifting environments further suggests that transient polygenic adaptation may be widespread[6,71,98]. Just how common we should expect such adaptive bouts to be depends, among other factors, on the time scale over which environments change (relative to the time scale of causal allele dynamics) and their degree of autocorrelation across time[52,99], and remains to be investigated. In any case, these considerations raise interesting questions about the proper framing of tests for polygenic adaptation, if indeed polygenic adaptation has been the norm rather than the exception in human evolution.

**Gene by environment interactions**

In discussing how traits are expected to evolve, we have ignored the possibility of gene by environment (GxE) interactions, in which the genetic effects manifest themselves differently in distinct environments. In model organisms, in which environments can be controlled and manipulated, such interactions are ubiquitous (e.g., refs. 2,100,101). In humans, the same types of experiments obviously cannot be conducted, and statistical approaches to detect GxE have identified few clear-cut cases[25]. If GxE interactions turn out to be prevalent in humans too, they would only amplify the points highlighted in this Perspective: (1) that we should expect polygenic effects to differ across groups and (2) that relating differences in polygenic effects to trait differences requires strong assumptions about environmental effects.

GxE interactions are more than a complication for the models whose results we have summarized, however. Their existence challenges the validity of the generative models on which we rely to make inferences about complex traits (e.g., heritability estimation[102]). These models envision variation in a trait as arising from strictly genetic effects and independent environmental effects, then sometimes consider a specific way in which genetics and environment could interact. For traits such as behaviors, for which genetic effects plausibly arise entirely and inextricably by interaction with environmental effects, it is not obvious to us that this conception is apt—even when a statistical model based on it appears to capture substantial trait variance[29,40,103,104]. If instead it is more sensible to think of such traits as emerging from GxE interactions, polygenic effects will be incommensurable across environments. These considerations are beyond the scope of this Perspective, but seem to us worth keeping in mind, especially when we then rely on these models to interpret differences among people over time and across the globe.

## Conclusion

Whether a trait is neutral, its fitness optimum recently changed or its fitness optimum has remained the same in the face of shifting environmental effects, we should expect mean polygenic effects to evolve. Therefore, a shift in polygenic effects does not imply that mean trait values have changed. Nor does it follow without strong assumptions about environmental effects that if mean trait values have changed, they will mirror the direction or magnitude of the genetic changes. Such considerations apply to any organism (e.g., refs. 99,105), but are particularly acute for humans and other non-experimental species, for which we almost always lack the ability to render environmental effects in different groups identical, as done in common garden experiments for example. As a consequence, there may be little that can be reliably inferred from comparisons of trait genetics across human groups, even when current limitations of PGS are overcome. In the absence of knowledge of the salient environmental effects (and perhaps their interaction with genetics), a fundamental interpretative ambiguity will remain.

**Acknowledgments**. We thank Ipsita Agarwal, Dalton Conley, Graham Coop, Anna Di Rienzo, Doc Edge, Michael B. Eisen, Laura Hayward, Magnus Nordborg, Jonathan Pritchard, Bahareh Rashidi, Guy Sella and Eric Turkheimer for helpful discussions and/or comments on an earlier version of the manuscript. The work was supported by a fellowship from the Simons Foundation's Society of Fellows (#633313) to AH and NIH GM121372 to MP.

**Literature cited**

1. Lynch, M. & Walsh, B. *Genetics and analysis of quantitative traits*. (Sinauer Sunderland, MA, 1998).
2. Falconer, D. S. & Mackay, T. F. C. *Introduction to quantitative genetics*. (Pearson Education India, 1996).
3. Rockman, M. V. The QTN program and the alleles that matter for evolution: All that's gold does not glitter. *Evolution (N. Y).* **66**, 1–17 (2012).
4. Visscher, P. M., Brown, M. A., McCarthy, M. I. & Yang, J. Five years of GWAS discovery. *Am. J. Hum. Genet.* **90**, 7–24 (2012).
5. Claussnitzer, M. *et al.* A brief history of human disease genetics. *Nature* **577**, 179–189 (2020).
6. Sella, G. & Barton, N. H. Thinking About the Evolution of Complex Traits in the Era of Genome-Wide Association Studies. *Annu. Rev. Genomics Hum. Genet.* **20**, 461–493 (2019).
7. Fisher, R. A. *The genetical theory of natural selection*. (Clarendon Press, 1930).
8. Yengo, L. *et al.* Meta-analysis of genome-wide association studies for height and body mass index in ~700000 individuals of European ancestry. *Hum. Mol. Genet.* **27**, 3641–3649 (2018).
9. Wray, N. R., Kemper, K. E., Hayes, B. J., Goddard, M. E. & Visscher, P. M. Complex trait prediction from genome data: contrasting EBV in livestock to PRS in humans: genomic prediction. *Genetics* **211**, 1131–1141 (2019).
10. Choi, S. W., Mak, T. S.-H. & O'Reilly, P. F. Tutorial: a guide to performing polygenic risk score analyses. *Nat. Protoc.* 1–14 (2020).
11. Power, R. A. *et al.* Polygenic risk scores for schizophrenia and bipolar disorder predict creativity. *Nat. Neurosci.* **18**, 953–955 (2015).
12. Khera, A. V. *et al.* Genome-wide polygenic scores for common diseases identify individuals with risk equivalent to monogenic mutations. *Nat. Genet.* **50**, 1219–1224 (2018).
13. Torkamani, A., Wineinger, N. E. & Topol, E. J. The personal and clinical utility of polygenic risk scores. *Nat. Rev. Genet.* **19**, 581 (2018).
14. Mavaddat, N. *et al.* Polygenic Risk Scores for Prediction of Breast Cancer and Breast Cancer Subtypes. *Am. J. Hum. Genet.* **104**, 21–34 (2019).
15. Chen, R. *et al.* Type 2 Diabetes Risk Alleles Demonstrate Extreme Directional Differentiation among Human Populations , Compared to Other Diseases. **8**, (2012).
16. Martin, A. R. *et al.* Human Demographic History Impacts Genetic Risk Prediction across Diverse Populations. *Am. J. Hum. Genet.* **100**, 635–649 (2017).
17. Duncan, L. *et al.* Analysis of polygenic score usage and performance across diverse human populations. *BioRxiv* 398396 (2018).
18. Robinson, M. R. *et al.* Population genetic differentiation of height and body mass index across Europe. *Nat. Genet.* **47**, 1357 (2015).
19. Berg, J. J. & Coop, G. A population genetic signal of polygenic adaptation. *PLoS Genet.* **10**, e1004412 (2014).
20. Racimo, F., Berg, J. J. & Pickrell, J. K. Detecting Polygenic Adaptation in Admixture Graphs. *Genetics* **208**, 1565 LP – 1584 (2018).
21. Martin, A. R. *et al.* Clinical use of current polygenic risk scores may exacerbate health disparities. *Nat. Genet.* **51**, 441261 (2019).
22. Mostafavi, H. *et al.* Variable prediction accuracy of polygenic scores within an ancestry group. *Elife* **9**, e48376 (2020).
23. Berg, J. J. *et al.* Reduced signal for polygenic adaptation of height in UK biobank. *Elife* **8**, (2019).

24. Sohail, M. *et al.* Polygenic adaptation on height is overestimated due to uncorrected stratification in genome-wide association studies. *Elife* **8**, e39702 (2019).
25. Young, A. I., Benonisdottir, S., Przeworski, M. & Kong, A. Deconstructing the sources of genotype-phenotype associations in humans. *Science (80-. ).* **365**, 1396–1400 (2019).
26. Wang, Y. *et al.* Theoretical and empirical quantification of the accuracy of polygenic scores in ancestry divergent populations. *bioRxiv* (2020).
27. Duncan, L. *et al.* Analysis of polygenic risk score usage and performance in diverse human populations. *Nat. Commun.* **10**, 1–9 (2019).
28. Barton, N. H., Hermisson, J. & Nordborg, M. Population Genetics: Why structure matters. *Elife* **8**, e45380 (2019).
29. Coop, G. Reading tea leaves? Polygenic scores and differences in traits among groups. *arXiv Prepr. arXiv1909.00892* (2019).
30. Rosenberg, N. A., Edge, M. D., Pritchard, J. K. & Feldman, M. W. Interpreting polygenic scores, polygenic adaptation, and human phenotypic differences. *Evol. Med. public Heal.* **2019**, 26–34 (2018).
31. Barton, N. H. & Keightley, P. D. Understanding quantitative genetic variation. *Nat. Rev. Genet.* **3**, 11–21 (2002).
32. Turchin, M. C. *et al.* Evidence of widespread selection on standing variation in Europe at height-associated SNPs. *Nat. Genet.* **44**, 1015 (2012).
33. Thornton, K. R. Polygenic adaptation to an environmental shift: temporal dynamics of variation under Gaussian stabilizing selection and additive effects on a single trait. *Genetics* **213**, 1513–1530 (2019).
34. Hoellinger, I., Pennings, P. S. & Hermisson, J. Polygenic Adaptation: From sweeps to subtle frequency shifts. *PLoS Genet.* **15**, e1008035 (2019).
35. Barghi, N., Hermisson, J. & Schlötterer, C. Polygenic adaptation: a unifying framework to understand positive selection. *Nat. Rev. Genet.* 1–13 (2020).
36. de Vladar, H. P. & Barton, N. Stability and response of polygenic traits to stabilizing selection and mutation. *Genetics* **197**, 749–767 (2014).
37. Jain, K. & Stephan, W. Modes of rapid polygenic adaptation. *Mol. Biol. Evol.* **34**, 3169–3175 (2017).
38. Hayward, L. K. & Sella, G. Polygenic adaptation after a sudden change in environment. *BioRxiv* 792952 (2019).
39. Edge, M. D. & Rosenberg, N. A. A general model of the relationship between the apportionment of human genetic diversity and the apportionment of human phenotypic diversity. *Hum. Biol.* **87**, 313–337 (2015).
40. Novembre, J. & Barton, N. H. Tread lightly interpreting polygenic tests of selection. *Genetics* **208**, 1351 (2018).
41. Reich, D. *Who we are and how we got here: Ancient DNA and the new science of the human past*. (Oxford University Press, 2018).
42. Hill, W. G., Goddard, M. E. & Visscher, P. M. Data and theory point to mainly additive genetic variance for complex traits. *PLoS Genet* **4**, e1000008 (2008).
43. Zhu, Z. *et al.* Dominance genetic variation contributes little to the missing heritability for human complex traits. *Am. J. Hum. Genet.* **96**, 377–385 (2015).
44. Simons, Y. B., Bullaughey, K., Hudson, R. R. & Sella, G. A population genetic interpretation of GWAS findings for human quantitative traits. *PLoS Biol.* **16**, e2002985 (2018).
45. Höllinger, I., Pennings, P. S. & Hermisson, J. Polygenic adaptation: From sweeps to subtle frequency shifts. *PLoS Genet.* **15**, e1008035 (2019).
46. Whitlock, M. C. & Guillaume, F. Testing for spatially divergent selection: comparing QST to FST. *Genetics* **183**, 1055–1063 (2009).
47. Leinonen, T., McCairns, R. J. S., O'hara, R. B. & Merilä, J. Q ST--F ST comparisons:

evolutionary and ecological insights from genomic heterogeneity. *Nat. Rev. Genet.* **14**, 179–190 (2013).
48. Barton, N. H. Pleiotropic models of quantitative variation. *Genetics* **124**, 773–782 (1990).
49. Boyle, E. A., Li, Y. I. & Pritchard, J. K. An expanded view of complex traits: from polygenic to omnigenic. *Cell* **169**, 1177–1186 (2017).
50. Pickrell, J. K. *et al.* Detection and interpretation of shared genetic influences on 42 human traits. *Nat. Genet.* **48**, 709 (2016).
51. Watanabe, K. *et al.* A global overview of pleiotropy and genetic architecture in complex traits. *Nat. Genet.* **51**, 1339–1348 (2019).
52. Stearns, S. C., Byars, S. G., Govindaraju, D. R. & Ewbank, D. Measuring selection in contemporary human populations. *Nat. Rev. Genet.* **11**, 611–622 (2010).
53. Pritchard, J. K. & Di Rienzo, A. Adaptation--not by sweeps alone. *Nat. Rev. Genet.* **11**, 665 (2010).
54. Fan, S., Hansen, M. E. B., Lo, Y. & Tishkoff, S. A. Going global by adapting local: A review of recent human adaptation. *Science (80-. ).* **354**, 54–59 (2016).
55. Mathieson, I. Human adaptation over the past 40,000 years. *Curr. Opin. Genet. Dev.* **62**, 97–104 (2020).
56. Nielsen, R. *et al.* Tracing the peopling of the world through genomics. *Nature* **541**, 302–310 (2017).
57. Jablonski, N. G. & Chaplin, G. Human skin pigmentation as an adaptation to UV radiation. *Proc. Natl. Acad. Sci.* **107**, 8962–8968 (2010).
58. Balding, D. J., Bishop, M. & Cannings, C. *Handbook of statistical genetics*. (John Wiley & Sons, 2008).
59. Gilad, Y., Oshlack, A., Smyth, G. K., Speed, T. P. & White, K. P. Expression profiling in primates reveals a rapid evolution of human transcription factors. *Nature* **440**, 242–245 (2006).
60. Sanjak, J. S., Sidorenko, J., Robinson, M. R., Thornton, K. R. & Visscher, P. M. Evidence of directional and stabilizing selection in contemporary humans. *Proc. Natl. Acad. Sci.* **115**, 151–156 (2018).
61. Gibson, G. Decanalization and the origin of complex disease. *Nat. Rev. Genet.* **10**, 134–140 (2009).
62. Gibson, G. & Lacek, K. A. Canalization and Robustness in Human Genetics and Disease. *Annu. Rev. Genet.* **54**, (2020).
63. Karn, M. N. & Penrose, L. S. Birth weight and gestation time in relation to maternal age, parity and infant survival. *Ann. Eugen.* **16**, 147–164 (1951).
64. Hill, W. G. & Keightley, P. D. Interrelations of mutation, population size, artificial and natural selection. in *Proceedings of the second international conference on quantitative genetics* 57–70 (1988).
65. Edge, M. D. & Coop, G. Reconstructing the history of polygenic scores using coalescent trees. *Genetics* **211**, 235–262 (2019).
66. Pak, S., Schwekendiek, D. & Kim, H. K. Height and living standards in North Korea, 1930s--1980s. *Econ. Hist. Rev.* **64**, 142–158 (2011).
67. Selzam, S. *et al.* Comparing within- and between-family polygenic score prediction Authors. *bioRxiv* 1–32 (2019).
68. Ge, T., Chen, C.-Y., Neale, B. M., Sabuncu, M. R. & Smoller, J. W. Phenome-wide heritability analysis of the UK Biobank. *PLoS Genet.* **13**, e1006711 (2017).
69. Tyrrell, J. *et al.* Genetic predictors of participation in optional components of UK Biobank. *bioRxiv* (2020).
70. Levins, R. Thermal acclimation and heat resistance in Drosophila species. *Am. Nat.* **103**, 483–499 (1969).
71. Conover, D. 0. & Schultz, E. T. Phenotypic similarity and the evolutionary significance of

counter­gradient variation. *Trends Ecol. Evol.* **10**, 248–252 (1995).
72. Grether, G. F. Environmental change, phenotypic plasticity, and genetic compensation. *Am. Nat.* **166**, E115--E123 (2005).
73. Storz, J. F., Scott, G. R. & Cheviron, Z. A. Phenotypic plasticity and genetic adaptation to high-altitude hypoxia in vertebrates. *J. Exp. Biol.* **213**, 4125–4136 (2010).
74. Turelli, M. Heritable genetic variation via mutation-selection balance: Lerch's zeta meets the abdominal bristle. *Theor. Popul. Biol.* **25**, 138–193 (1984).
75. de Visser, J. A. G. M. *et al.* Perspective: Evolution and Detection of Genetic Robustness. *Evolution (N. Y).* **57**, 1959–1972 (2003).
76. Hill, W. G. & Mulder, H. A. Genetic analysis of environmental variation. *Genet. Res. (Camb).* **92**, 381–395 (2010).
77. Paaby, A. B. & Rockman, M. V. Cryptic genetic variation: Evolution's hidden substrate. *Nat. Rev. Genet.* **15**, 247–258 (2014).
78. Charlesworth, B. Some quantitative methods for studying evolutionary patterns in single characters. *Paleobiology* 308–318 (1984).
79. Urban, M. C. *et al.* Evolutionary origins for ecological patterns in space. *Proc. Natl. Acad. Sci. U. S. A.* **117**, 17482–17490 (2020).
80. Martin, A. R. *et al.* Human Demographic History Impacts Genetic Risk Prediction across Diverse Populations. *Am. J. Hum. Genet.* **107**, 788–789 (2020).
81. Kerminen, S. *et al.* Geographic Variation and Bias in the Polygenic Scores of Complex Diseases and Traits in Finland. *Am. J. Hum. Genet.* **104**, 1169–1181 (2019).
82. Coop, G. *et al.* The role of geography in human adaptation. *PLoS Genet.* **5**, (2009).
83. Yang, H. C., Lee, L. K. & Co, R. S. A low jitter 0.3-165 MHz CMOS PLL frequency synthesizer for 3 V/5 V operation. *IEEE J. Solid-State Circuits* **32**, 582–586 (1997).
84. Hernandez, R. D. *et al.* Classic Selective Sweeps Were Rare in Recent Human Evolution. *Science (80-. ).* **331**, 920 LP – 924 (2011).
85. Field, Y. *et al.* Detection of human adaptation during the past 2000 years. *Science (80-. ).* **354**, 760–764 (2016).
86. Uricchio, L. H., Kitano, H. C., Gusev, A. & Zaitlen, N. A. An evolutionary compass for detecting signals of polygenic selection and mutational bias. *Evol. Lett.* **3**, 69–79 (2019).
87. Speidel, L., Forest, M., Shi, S. & Myers, S. A method for genome-wide genealogy estimation for thousands of samples. *BioRxiv* 550558 (2019).
88. Stern, A. J., Speidel, L., Zaitlen, N. A. & Nielsen, R. Disentangling selection on genetically correlated polygenic traits using whole-genome genealogies. *bioRxiv* (2020).
89. Zeng, J. *et al.* Signatures of negative selection in the genetic architecture of human complex traits. *Nat. Genet.* **50**, 746–753 (2018).
90. Lander, E. S. & Schork, N. J. Genetic dissection of complex traits. *Nat. Genet.* **12**, 355–356 (1996).
91. Haworth, S. *et al.* Apparent latent structure within the UK Biobank sample has implications for epidemiological analysis. *Nat. Commun.* **10**, 333 (2019).
92. Lande, R. & Arnold, S. J. The Measurement of Selection on Correlated Characters. *Evolution (N. Y).* **37**, 1210–1226 (1983).
93. Berg, J. J., Zhang, X. & Coop, G. Polygenic Adaptation has Impacted Multiple Anthropometric Traits. (2019).
94. Ruff, C. B. Morphological adaptation to climate in modern and fossil hominids. *Am. J. Phys. Anthropol.* **37**, 65–107 (1994).
95. Foster, F. & Collard, M. A reassessment of Bergmann's rule in modern humans. *PLoS One* **8**, e72269 (2013).
96. Crawford, N. G. *et al.* Loci associated with skin pigmentation identified in African populations. *Science (80-. ).* **358**, eaan8433 (2017).
97. Mathieson, I. *et al.* Genome-wide patterns of selection in 230 ancient Eurasians. *Nature*

(2015). doi:10.1038/nature16152
98. Conover, D. O., Duffy, T. A. & Hice, L. A. The covariance between genetic and environmental influences across ecological gradients: reassessing the evolutionary significance of countergradient and cogradient variation. *Ann. N. Y. Acad. Sci.* **1168**, 100–129 (2009).
99. Pemberton, J. M. Evolution of quantitative traits in the wild : mind the ecology. 2431–2438 (2010). doi:10.1098/rstb.2010.0108
100. Des Marais, D. L., Hernandez, K. M. & Juenger, T. E. Genotype-by-environment interaction and plasticity: Exploring genomic responses of plants to the abiotic environment. *Annu. Rev. Ecol. Evol. Syst.* **44**, 5–29 (2013).
101. Falconer, D. S. The Problem of Environment and Selection. *Am. Nat.* **86**, 293–298 (1952).
102. Vinkhuyzen, A. A. E., Wray, N. R., Yang, J., Goddard, M. E. & Visscher, P. M. Estimation and partition of heritability in human populations using whole-genome analysis methods. *Annu. Rev. Genet.* **47**, 75–95 (2013).
103. Turkheimer, E. Three laws of behavior genetics and what they mean. *Curr. Dir. Psychol. Sci.* **9**, 160–164 (2000).
104. Lewontin, R. C. Annotation: the analysis of variance and the analysis of causes. *Am. J. Hum. Genet.* **26**, 400 (1974).
105. Kawecki, T. J. & Ebert, D. Conceptual issues in local adaptation. *Ecol. Lett.* **7**, 1225–1241 (2004).